\documentclass[
    ,final            
  ]
  {aipproc}

\layoutstyle{6x9}

\begin{document}

\title{Consequences of Leading-Logarithm Summation for the Radiative Breakdown of Standard-Model Electroweak Symmetry}

\author{V. Elias}{
  address={Department of Applied Mathematics, The University of Western Ontario, \\ London, Ontario N6A 5B7, Canada}
  ,altaddress={Perimeter Institute for Theoretical Physics, 35 King Street North, Waterloo, ON  N2J 2W9, Canada}
}

\author{R. B. Mann}{
  address={Perimeter Institute for Theoretical Physics, 35 King Street North, Waterloo, ON N2J 2W9, Canada},
  ,altaddress={Department of Physics, University of Waterloo, Waterloo, ON N2L 3G1, Canada}
}

\author{D. G. C. McKeon}{
  address={Department of Applied Mathematics, The University of Western Ontario, \\ London, Ontario N6A 5B7, Canada}
}

\author{T. G. Steele}{
  address={Department of Physics and Engineering Physics, University of Saskatchewan, \\ Saskatoon, Saskatchewan S7N 5E2, Canada}
}

\begin{abstract}
In the empirically sensible limit in which QCD, $t$-quark Yukawa, and scalar-field-interaction coupling constants dominate all other Standard-Model coupling constants, we sum all leading-logarithm terms within the perturbative expansion for the effective potential that contribute to the extraction of the Higgs boson mass via radiative electroweak symmetry breaking. A Higgs boson mass of $216 \; GeV$ emerges from such terms, as well as a scalar-field-interaction coupling constant substantially larger than that anticipated from conventional spontaneous symmetry breaking. The sum of the effective potential's leading logarithms is shown to exhibit a local minimum in the limit $\phi \rightarrow 0$ if the QCD coupling constant is sufficiently strong, suggesting (in a multiphase scenario) that electroweak physics may provide the mechanism for choosing the asymptotically-free phase of QCD.
\end{abstract}

\maketitle


\section{1. Radiative Electroweak Symmetry Breaking}

Radiative symmetry breaking, in which a vacuum expectation value arises from radiative corrections to a potential with no quadratic mass term, was first addressed in a classic paper by S. Coleman and E. Weinberg \cite{1}. Unlike conventional symmetry breaking, in which an arbitrary but negative mass term leads to a correspondingly arbitrary Higgs boson mass, the radiative scenario for electroweak symmetry breaking necessarily predicts the Higgs boson mass as well as the magnitude of the quartic scalar-field interaction coupling constant $\lambda$. Unfortunately, the first such predictions preceded the discovery of the top-quark, whose large Yukawa coupling dominates all one-loop radiative effects. Indeed, the large magnitude of this coupling constant destroys not only the applicability of Coleman and Weinberg's prediction for the Higgs mass (which is far below present empirical lower bounds), but also the use of a purely one-loop potential to make \emph{any} such predictions via radiative symmetry breaking \cite{2}. The only hope for the program of radiative symmetry breaking for electroweak physics is to include radiative effects past one-loop order. In the present work, we demonstrate how renormalization-group methods permit one to extract \emph{all} contributing leading-logarithm contributions to the Higgs boson mass, as well as some surprising results from the summation of leading logarithms in the zero field limit.

\section{2.  Leading Logarithms and the Higgs Boson Mass}

The dominant couplants of the single-Higgs-doublet standard model are the $t$-quark Yukawa couplant $\left( x \equiv g_t^2(v)/4\pi^2 = 2m_t^2/(2\pi v)^2 = 0.0253\right)$, the QCD gauge couplant $\left( z \equiv \alpha_s (v) /4\pi^2 = 0.0329\right)$ as evolved from $\alpha_s (M_Z) = 0.12$ \cite{3}, and the unknown quartic scalar couplant $y \equiv \lambda(v) / 4\pi^2$, where $v = 2^{-1/4} G_F^{-1/2} = 246 \; GeV$ is the expectation value characterizing the breakdown of electroweak symmetry. The remaining Yukawa and gauge couplants are all small compared to these three couplants, and are therefore ignored in the treatment which follows.

The effective potential of an $SU(2)\times U(1)$ single-Higgs-doublet $(\phi)$ theory in which a quadratic mass term is absent $\left( V_{tree} = \lambda(\phi^\dag \phi)^2 / 4 \equiv \lambda \phi^4 / 4 \right)$ satisfies the renormalization-group (RG) equation

\begin{eqnarray}
0 & = & \mu \frac{d}{d\mu} V \left[ \lambda(\mu), g_t (\mu), g_3 (\mu), \phi^2
(\mu), \mu \right] \nonumber\\
& = & \left( \mu \frac{\partial}{\partial\mu} + \beta_\lambda \frac{\partial}{%
\partial \lambda} + \beta_t \frac{\partial}{\partial g_t} + \beta_3 \frac{%
\partial}{\partial g_3} - 2\gamma \phi^2 \frac{\partial}{\partial \phi^2}
\right) V \left( \lambda, g_t, g_3, \phi^2, \mu \right),
\end{eqnarray}
This potential may be expressed in the form
\begin{equation}
V = \pi^2 \phi^4 \sum_{n=0}^\infty x^n \sum_{k=0}^\infty y^k \sum_{\ell = 0}^\infty z^\ell \sum_{p=0}^{n+k+\ell - 1} L^p \; D_{n,k,\ell,p}, \; [D_{0,1,0,0} = 1, \; D_{1,0,0,0} = D_{0,0,1,0} = 0],
\end{equation}
with the logarithm $L \equiv \log (\phi^2 / \mu^2)$ referenced to an arbitrary renormalization scale $\mu$.  The leading logarithm contributions to this potential arise when $p = n + k + \ell - 1$, i.e., when the degree of the logarithm is only one less than the sum of the powers of the three contributing couplants.  If we define $C_{n,k,\ell} \equiv D_{n,k,\ell, n + k + \ell - 1}$, such coefficients of the leading logarithm series,
\begin{eqnarray}
V_{LL} & = & \pi^2 \phi^4 S_{LL} = \pi^2 \phi^4 \left\{\sum_{n=0}^\infty x^n
\sum_{k=0}^\infty y^k \sum_{\ell=0}^\infty z^{\ell} C_{n, k, \ell}
L^{n+k+\ell-1} \right\},\nonumber\\ 
&& C_{0,0,0} = C_{1,0,0} = C_{0,0,1} = 0, \; \; C_{0,1,0} = 1, 
\end{eqnarray}
may be obtained via the explicit one-loop RG functions \cite{2,4} appearing in Eq. (1):
\begin{equation}
\left[ -2 \frac{\partial}{\partial L} + \left( \frac{9}{4} x^2 - 4xz \right) 
\frac{\partial}{\partial x} + \left( 6y^2 + 3yx - \frac{3}{2} x^2 \right) 
\frac{\partial}{\partial y} -\frac{7}{2} z^2 \frac{\partial}{\partial z} -
3x \right] S_{LL} (x,y,z,L) = 0.
\end{equation}
One can solve this equation for successive powers of $L$.  For example, the aggregate coefficient of $L^0$ is given by
\begin{equation}
- 2 \left( C_{0,2,0} \; y^2 + C_{2,0,0} \; x^2 + C_{0,0,2} \; z^2 +
C_{1,1,0} \; xy + C_{1,0,1} \; xz + C_{0,1,1} yz \right) + 6y^2 - \frac{3}{2}
x^2 = 0,
\end{equation}
in which case $C_{0,2,0} = 3$, $C_{2,0,0} = -\frac{3}{4}$, and the remaining degree-2 coefficients within Eq. (3) are zero:
\begin{equation}
S_{LL} = y + 3y^2 L - \frac{3}{4} x^2 L + \ldots = \frac{\lambda}{4\pi^2} +
\left( \frac{3\lambda^2}{16\pi^4} - \frac{3g_t^4}{64\pi^4} \right) \log
\left( \frac{\phi^2}{\mu^2} \right) + \ldots  
\end{equation}
Eq. (6) corresponds to the ${\cal{O}}(\lambda^2, g_t^2)$ diagrammatic contributions to the $SU(2)  \times U(1)$ effective potential $\left( V_{LL} = \pi^2 \phi^4 S_{LL} \right)$ calculated in ref. \cite{1}.  Such a  brute-force approach can be utilized to obtain all subsequent $C_{n,k,\ell}$ coefficients.  The extraction of a Higgs boson mass, however, is sensitive only to terms in $S_{LL}$ of degree-4 or less in $L$.  These terms are given by
\begin{equation}
S_{LL} = y + BL + CL^2 + DL^3 + EL^4 + \ldots
\end{equation}
where \cite{5}
\begin{equation}
B = 3y^2 - \frac{3}{4} x^2,
\end{equation}
\begin{equation}
C = 9y^3 + \frac{9}{4} xy^2 - \frac{9}{4} x^2 y + \frac{3}{2} x^2 z - \frac{9%
}{32} x^3, 
\end{equation}
\begin{equation}
D = 27 y^4 + \frac{27}{2} x y^3 - \frac{3}{2} xy^2 z + 3x^2 yz - \frac{225}{%
32} x^2y^2 - \frac{23}{8} x^2z^2 + \frac{15}{16} x^3z - \frac{45}{16} x^3 y
+ \frac{99}{256} x^4,
\end{equation}
\begin{eqnarray}
E & = & 81y^5 + \frac{243}{4} xy^4 - 9xy^3z + \frac{45}{32} x y^2 z^2 - \frac{%
69}{16} x^2 yz^2 - \frac{135}{8} x^2 y^3 + \frac{531}{64} x^2 y^2 z \nonumber\\
& + & \frac{345}{64} x^2 z^3 - \frac{603}{256} x^3 z^2 + \frac{207}{32} x^3 yz
- \frac{8343}{512} x^3 y^2 - \frac{459}{512} x^4 z + \frac{135}{512} x^4 y + 
\frac{837}{1024} x^5.\nonumber\\
\end{eqnarray}

We emphasize that Eqs. (7) - (11) represent the sum of \emph{all} leading-logarithm terms contributing to the Higgs boson mass in the radiatively-broken single-Higgs-doublet standard model.  The extraction of a mass proceeds in the same manner as described in ref. \cite{1} for the one-loop potential.  After subtractions, $V$ will contain a finite $K \phi^4$ counterterm,
\begin{equation}
V = \pi^2 \phi^4 (S_{LL} + K),
\end{equation}
whose magnitude is determined by renormalization conditions.  Formally this counterterm may be identified with the sum of non-leading $p = 0$ contributions to Eq. (2),
\begin{equation}
K = \sum_{n=0}^\infty \sum_{k=0}^\infty \sum_{\ell = 0}^\infty x^n y^k z^\ell D_{n,\ell, k, 0}, \; \;\; n+k+\ell \geq 2.
\end{equation}
This sum represents a combination of non-leading logarithm contributions to $V$, a sum of terms in Eq. (2) whose couplants have an aggregate power at least two larger than the power (zero) of the logarithm.  Consequently, the counterterm $K$ is not RG-accessible via Eq. (4), but must be determined by a set of renormalization conditions.  In the complete absence of a bare $\phi^2$ mass term, note also that the potential will not generate a renormalized $\phi^2$ term;  the (external-to-Standard-Model) symmetry that would permit a conformally invariant tree-potential \cite{2} will be preserved within the context of a gauge-invariant regularization procedure \cite{6}.

The set of renormalization conditions we employ at the choice $\mu = v$ for renormalization scale are the same as in ref. \cite{1}:
\begin{equation}
\left. \frac{dV}{d\phi}\right|_v =0, \; \; \left. \frac{d^2 V}{d\phi^2}\right|_v = m_\phi^2, \; \; \left. \frac{d^4 V}{d \phi^4}\right|_v = \frac{d^4}{d\phi^4} \left( \frac{\lambda \phi^4}{4} \right) = 24\pi^2 y
\end{equation}
which respectively imply that
\begin{equation}
4(y + K) + 2B = 0
\end{equation}
\begin{equation}
\pi^2 v^2 (12y + 12K + 14B + 8C) = m_\phi^2
\end{equation}
\begin{equation}
24(y+K) + 100B + 280C + 480D + 384E = 0.
\end{equation}
Note from Eqs. (8) - (11) that $B$, $C$, $D$ and $E$ are all functions of the undetermined couplant $y$, as $x$ and $z$ are known empirically.  Consequently, the factor of $y+K$ can be eliminated between Eqs. (15) and (17) to obtain a fifth-order equation for $y$ with three real solutions, $\left\{ 0.0538, -0.278, -0.00143 \right\}$.  For a given choice of solution, the coefficients $\left\{ B, C, D, E \right\}$ are numerically determined.  One finds from Eqs. (15) and (16) that the respective values for the square of the Higgs mass $(=8\pi^2 v^2 (B+C))$ are positive only for the first two choices for $y$, of which only the first has any likelihood of perturbative stability (see below);  we thus find for $y = 0.0538$ that $m_\phi = 216 \; GeV$.  

\section{3.  Phenomenological Viability and Consequences}

The estimate $m_\phi = 216 \; GeV$ is much larger than the ${\cal{O}}(10 \; GeV)$ estimate obtained from radiative symmetry breaking \cite{1} in the absence of any empirical knowledge of the $t$-quark.  The $216 \; GeV$ estimate is also within striking distance of the indirect 95\% confidence-level phenomenological bound $m_\phi \leq 196 \; GeV$ \cite{3} obtained from $\log(m_\phi)$ factors in radiative corrections to $m_W$, $m_Z$ and $\Gamma_Z$.  However, the viability of this estimate rests upon its stability under subsequent (next-to-leading logarithm) corrections, which we are not yet able to compute.

If residual renormalization-scale dependence of the potential is indicative of the magnitude of such subsequent corrections, we have found that the extimate $m_\phi = 216 \; GeV$ is quite stable.  If we allow $\mu$ to vary from $v/2$ to $2v$ within leading-logarithm (but not counterterm \footnote{We assume $K\phi^4$ to be RG-invariant in the one-loop sense, since the counterterm corresponds to a sum of subsequent-to-leading logarithm contributions, as noted above.}) contributions to the potential, with concomitant evolution of $x(\mu)$, $y(\mu)$ and $z(\mu)$ from known values at $\mu = v$, and $\phi(\mu)$ from its input value $\phi(\equiv \phi(v))$, 
we find the corresponding range of values for $m_\phi$ to vary between $208$ and $217 \; GeV$ \cite{5}.  Although the $y = 0.0538$ value we obtain from Eqs. (15) and (17) is much larger than the quartic scalar-interaction couplant $(\lambda/4\pi^2)$ that would arise from generating a $216 \; GeV$ Higgs mass via conventional (non-radiative) symmetry breaking, this large couplant still appears to be perturbative.  In the limit $y >> x,z$, the scalar-field sector of the Standard Model decouples into an $O(4)$-symmetric scalar field theory whose $\beta$- and $\gamma$-functions are known to five-loop order \cite{7}:
\begin{equation}
\lim_{\stackrel{_{x \rightarrow 0}}{_{z \rightarrow 0}}} \mu \frac{dy}{d\mu}
= 6y^2 - \frac{39}{2} y^3 + 187.85y^4 - 2698.3y^5 + 47975 y^6 + \ldots \; \; ,
\end{equation}
\begin{equation}
\lim_{\stackrel{_{x \rightarrow 0}}{_{z \rightarrow 0}}}
\gamma \;\; \sim \;\;  y^2 \left[ 1 - \frac{3}{2} y + \frac{195}{16} y^2 - 132.9 y^3 +
...\right].
\end{equation}
Both series above decrease monotonically when $y = 0.0538$, consistent (though barely so) with $y$ being sufficiently small to be perturbative.  

As noted above, the salient phenomenological signature of radiative (as opposed to conventional) electroweak symmetry breaking appears to be the pairing of an empirically viable Higgs mass with a large scalar-field interaction couplant.  In conventional symmetry breaking a $216 \; GeV$ Higgs corresponds to a quartic scalar couplant $y = m_\phi^2/(8\pi^2 v^2) = 0.0093$, as opposed to the value $0.0538$ obtained above.  Consequently, $y$-sensitive processes such as the longitudinal channel for $W^+ W^- \rightarrow ZZ$ scattering, in which $\sigma \sim y^2$ \cite{8}, should serve to distinguish between the two approaches to symmetry breaking, with \emph{an order of magnitude enhancement} predicted for the radiative case.  

\section{4.  Large Logarithm Behaviour}

A salient motivation for summing the leading logarithms of any given process is to ascertain the large logarithm limit of that process, since leading logarithm terms dominate all subsequent terms when the logarithm is itself large.  In the case of the Standard-Model effective potential, the large logarithm limit corresponds either to $|\phi| \rightarrow \infty$ or $\phi \rightarrow 0$ behaviour of the potential.  To extract this behaviour, we first express the leading-logarithm series $S_{LL}$ in the form
\begin{equation}
S_{LL} = y F_0 (w, \zeta) + \sum_{n=1}^\infty x^n L^{n-1} F_n (w, \zeta), \; \; \zeta \equiv zL, \; \; w \equiv 1 - 3yL.
\end{equation}
If we substitute Eq. (20) into the RG equation (4), we obtain the following recursive set of partial differential equations \cite{5}

\begin{equation}
\zeta \left( 1 + \frac{7}{4} \zeta \right) \frac{\partial}{\partial \zeta}
F_0 (w, \zeta) = (1 - w) \left[ w\frac{\partial}{\partial w} + 1 \right] F_0
(w, \zeta),
\end{equation}
\begin{equation}
\left[ \zeta \left( 1 + \frac{7}{4} \zeta \right) \frac{\partial}{\partial
\zeta} + 2\zeta + (w - 1) w \frac{\partial}{\partial w} \right] F_1 (w,
\zeta) = -\frac{(w-1)^2}{2} \frac{\partial}{\partial w} F_0 (w, \zeta),
\end{equation} 
\begin{eqnarray}
&& \left[ \zeta \left( 1 + \frac{7}{4} \zeta \right) \frac{\partial}{\partial
\zeta} + (w-1) w \frac{\partial}{\partial w} + (1 + 4\zeta) \right] F_2 (w,
\zeta)\nonumber\\
& = & \left[ \frac{3}{2} (w-1) \frac{\partial}{\partial w} - 
\frac{3}{8} \right] F_1 (w, \zeta) - \frac{3}{4} \left[ (w-1) \frac{\partial%
}{\partial w} + 1 \right] F_0 (w, \zeta),
\end{eqnarray}
\begin{eqnarray}
&& \left[ \zeta \left(1 + \frac{7}{4} \zeta \right) \frac{\partial}{\partial
\zeta} + (w-1) w \frac{\partial}{\partial w} + (k-1 + 2k \zeta) \right] F_k
(w, \zeta)\nonumber \\
& = & \left[ \frac{3(3k-7)}{8} + \frac{3}{2} (w-1) \frac{%
\partial}{\partial w} \right] F_{k-1} (w, \zeta) + \frac{9}{2} \frac{%
\partial F_{k- 2}}{\partial w} (w, \zeta), \; \; \; (k \geq 3).
\end{eqnarray}
Given that $F_0 (1,0) = 1$, i.e., that $S_{LL} \rightarrow y$ as $L \rightarrow 0$, one finds from Eq. (21) that
\begin{equation}
F_0 (w,\zeta) = 1/w.
\end{equation}
One then finds that the only solution to Eq. (22) for $F_1(w,\zeta)$ that is non-singular at $\zeta = 0$ [i.e. non-singular when QCD is turned off $(z = 0)$] is
\begin{eqnarray}
F_1 (w, \zeta) & = & \left[ \frac{6\zeta + 4[1-(1+7\zeta/4)^{6/7}]}{3\zeta^2}\right] \left( \frac{w-1}{w}\right)^2\nonumber\\
& = & \left[ \frac{1}{4} - \frac{\zeta}{6} + \frac{5}{32} \zeta^2 + \; \ldots \right]\left( \frac{w-1}{w} \right)^2 .
\end{eqnarray}

Solutions of the form
\begin{equation}
F_p(w,\zeta) = \sum_{k=0}^{p+1} f_{p,k} (\zeta) \left[ \frac{w-1}{w}\right]^k
\end{equation}
can be found by straightforward means from Eqs. (23) and (24).  For the $p = 2$ case, we find that
\begin{equation}
f_{2,0}(\zeta )=\left[ Z^{-9/7}-1\right] /3\zeta ,
\end{equation}
\begin{equation}
f_{2,1}(\zeta )=\left[ 2\zeta -4(1-Z^{-2/7})\right] /3\zeta ^{2},
\end{equation}
\begin{equation}
f_{2,2}(\zeta )=\left[ 20+71\zeta /2-\zeta ^{2}+22Z^{5/7}-42Z^{6/7}\right]
/3\zeta ^{3},
\end{equation}
\begin{equation}
f_{2,3}(\zeta )=\left[ -16-48\zeta -36\zeta
^{2}+32Z^{6/7}/7-16Z^{12/7}+192Z^{13/7}/7\right] /3\zeta ^{4},
\end{equation}
where $Z \equiv 1 + 7\zeta/4$.  For $p \geq 3$ we find from Eq. (24) that
\begin{eqnarray}
0 & = & \left( \left[ (7\zeta ^{2}/2)\frac{d}{d\zeta }+4p\zeta \right] +\left[
2\zeta \frac{d}{d\zeta }+2(p-1)+2k\right] \right) f_{p,k}(\zeta ) \nonumber\\
& - &\left[ (9p-21)/4+3k\right] f_{p-1,k}(\zeta )+3(k-1)f_{p-1,k-1}(\zeta ) \nonumber\\
& - & \left[ 9(k-1)/2\right] f_{p-2,k-1}(\zeta )+9kf_{p-2,k}(\zeta )-\left[
9(k+1)/2\right] f_{p-2,k+1}(\zeta ),
\end{eqnarray}
where $f_{p,k}\equiv 0$ when $k<0$ or $k>p+1$.

In the large $L$ limit, one finds from Eqs. (25) - (31) that the leading terms in the series (20) exhibit the following large-$L$ behaviour:
\begin{equation}
y \; F_0 \rightarrow - \frac{1}{3L}, \; \; \; x \; F_1 \rightarrow \frac{2}{L} \left( \frac{x}{z}\right), \; \; \; x^2 \; L F_2 \rightarrow -\frac{3}{2L} \left( \frac{x}{z} \right)^2.
\end{equation}
Moreover, since $(w - 1) / w \rightarrow 1$ when $|L| \rightarrow \infty$, we find that $F_n \rightarrow \sum_{k=0}^{n+1} f_{n,k} (\zeta)$.  We then find in the large $L$ (hence large-$\zeta$) limit of Eq. (32) that 
\begin{equation}
\left( \frac{7}{4} \zeta^2 \frac{d}{d\zeta} + 2n \zeta \right) F_n = \frac{%
3(3n-7)}{8} F_{n-1}, \; \; ( n \geq 3 ).  \label{eq8.17}
\end{equation}
a result that follows from the relations
\begin{equation}
\sum_{k=0}^{p+1} \left[ k f_{p-1, k} - (k-1) f_{p-1, k-1} \right] = 0,
\end{equation}
\begin{equation}
\sum_{k=0}^{p+1} \left[ - (k-1) f_{p-2, k-1} + 2k f_{p-2, k} - (k+1) f_{p-2, k+1}\right]=0,
\end{equation}
[Note $f_{p, k} = 0$ if $k > p+1$ or $k < 0$].
One then finds from Eqs. (33) and (34) that for $n > 2$,
\begin{equation}
F_n \rightarrow a_n \zeta^{-n}, \; \; a_n = \frac{3(3n-7)}{2n} a_{n-1}, \; \; a_2 = - 3/2.
\end{equation}
The results (33) and (37) permit explicit summation of the series (20) in the large $L$ limit, provided $x/z$ is sufficiently small \cite{5}:
\begin{equation}
V_{eff} \begin{array}{c}
{} \\ 
_{\longrightarrow} \\ 
^{|L| \rightarrow \infty}
\end{array}
\pi^2 \phi^4 S_{LL} \rightarrow - \frac{\pi^2 \phi^4}{3L} \left[1 - \frac{9x}{2z} \right]^{4/3}, \; \; \; 0 \leq x/z \leq 2/9.
\end{equation}

\section{5.  Footprints of New Physics}

If $x/z > 2/9$, the series (20) is outside its radius of convergence, and the result (38) is no longer applicable.  The result (38), therefore, is not relevant to empirical Standard-Model physics, for which $z = \alpha_s(v)/\pi = 0.033$, $x = 1.0/(4\pi^2)= 0.025$. However, this result \emph{is} of interest if QCD exhibits more than one phase.

Two phases for the evolution of the gauge coupling constant are known to characterize the exact $\beta$-function for N = 1 supersymmetric Yang-Mills theory in the absence of matter fields [9]. An asymptotically-free phase in which the gauge couplant is weak is accompanied by an additional non-asymptotically-free strong-couplant phase. Both weak and strong couplants evolve toward a common value at an infrared-attractive momentum scale, which serves as a lower bound on the domain of perturbative physics.

Pade approximant arguments have recently been advanced \cite{10} in support of QCD being characterized by similar two-phase behaviour. If such is the case, a very natural explanation emerges for the transition at $\mu \simeq m_\rho$ from QCD as a perturbative gauge theory of quarks and gluons to QCD as an effective theory of strongly-interacting hadrons \cite{10, 11}. However, within such a picture, there needs to be a mechanism for understanding why the ``weak'' asymptotically-free phase is the one we observe.

Eq. (38) shows an effective potential that approaches zero from above as $\phi \rightarrow 0$. Consequently, when the QCD couplant is sufficiently large, the effective potential exhibits a local \emph{minimum} at $\phi = 0$. Since the potential is itself zero at this minimum, whereas the empirical (weak-phase) potential is negative at its $\phi = v$ minimum, we see that the weaker of the two phases may be energetically-preferred by (radiatively-broken) electroweak symmetry.

As a final note, the large value for $y(v) = 0.0538$ emerging from radiative symmetry breaking suggests that the evolution of this couplant $[\mu(dy/d\mu) = 6y^2 + \ldots]$ is characterized by a Landau pole at (or below) $\mu = v \exp \{1/[6y(v)]\} \simeq 5.5 \; TeV$. Such a bound on the scale for new physics corresponds explicitly to the singularity at $w (= 1 - 3yL) = 0$ characterizing successive factors $x^p L^{p-1} F_p(w,\zeta)$ [Eq. (27)] within the leading-logarithm series (20). Thus, we can hopefully anticipate empirical evidence for the onset of new physics (or embedding symmetry) if electroweak symmetry is radiatively broken.


\begin{theacknowledgments}
We are grateful to the Natural Sciences and Engineering Research Council of Canada for support of this research.
\end{theacknowledgments}





\begin{thebibliography}{99}

\bibitem{1}S. Coleman and E. Weinberg,\emph{Phys. Rev.} {\bf D 7}, 1888 (1973)
\bibitem{2}M. Sher, {\em Phys. Rep.} {\bf 179}, 273 (1989)
\bibitem{3}Particle Data Group [K. Hagiwara et al.], {\em Phys. Rev.} {\bf D 66}, 1 (2002)
\bibitem{4}T. P. Cheng, E. Eichten and L. -F. Li, {\em Phys. Rev.} {\bf D 9}, 2259 (1974); M. B. Einhorn and D. R. T. Jones,  {\em Nucl. Phys.} {\bf B 211}, 29 (1983); M. J. Duncan, R. Philippe and M. Sher, {\em Phys. Lett.} {\bf B 153}, 165 (1985)
\bibitem{5}V. Elias, R. B. Mann, D. G. C. McKeon and T. G. Steele, hep-ph/0304153.
\bibitem{6}J. C. Collins, {\em Phys. Rev.} {\bf D 10}, 1213 (1974)
\bibitem{7}H. Kleinert, J. Neu, V. Schulte-Frohlinde, K. G. Chetyrkin and S. A. Larin, {\em Phys. Lett.} {\bf B 272}, 39 (1991);  {\bf B 319}, 545 (E) (1993)
\bibitem{8}U. Nierste and K. Riesselmann, {\em Phys. Rev.} {\bf D 53}, 6638 (1996)
\bibitem{9}I. I. Kogan and M. Shifman, \emph{Phys. Rev. Lett.} {\bf 75}, 2085 (1995); see also V. Elias, \emph{J. Phys.}{\bf G 27}, 217 (2001) regarding D. R. T. Jones, \emph{Phys. Lett.}{\bf B 123}, 45 (1983)
\bibitem{10}F. A. Chishtie, V. Elias, V. A. Miransky and T. G. Steele, \emph{Prog. Theor. Phys.} {\bf 104}, 603 (2000)
\bibitem{11}F. A. Chishtie, V. Elias and T. G. Steele, \emph{Phys. Lett.} {\bf B 514}, 279 (2001)
\end{thebibliography}
\end{document}